\documentclass[a4paper]{jpconf}
\usepackage{graphicx}

\usepackage{amsmath,mathrsfs,amssymb}
\usepackage{graphicx}

\begin{document}
\title{Fermion mixing with geometrical CP violation and its tests at the LHC}

\author{Ivo de Medeiros Varzielas}

\address{Fakult\"{a}t f\"{u}r Physik, Technische Universit\"{a}t
Dortmund D-44221 Dortmund, Germany}

\ead{ivo.de@udo.edu}


\begin{abstract}
We review and clarify some cases of geometrical CP violation, the framework of spontaneous CP violation through complex phases with values that are independent of parameters of the potential.
We present a flavour model based on $\Delta(27)$ featuring spontaneous CP violation, that can reproduce all quark masses and mixing data. The scalar sector of the model has exotic properties that can be tested at the LHC.
\end{abstract}


%


The work summarised here is mostly based on \,\cite{Varzielas:2012pd} and \,\cite{Bhattacharyya:2012pi}, which include more complete references that we omit here due to space constraints.

The origin of CP violation is an open question in particle physics. 
In the  Standard Model (SM), CP is violated through complex Yukawa couplings and this violation appears in charged weak interactions through the Cabibbo-Kobayashi-Maskawa (CKM) matrix.
In beyond the SM theories one can explore the origin of CP violation, and a very interesting framework is spontaneous CP violation: in such a framework,
CP is a symmetry of the Lagrangian and CP violation arises through complex vacuum expectation values (VEVs) of Higgs multiplets $H_i$. There is an important subtlety that should be noted. Even with complex VEVs, only when the unitary transformation $U$ acting on the $H_i$ and relating the VEV to its complex conjugate
\begin{equation}
\langle H_i \rangle \longrightarrow \langle H_{i} \rangle^{\ast} = U_{ij} \langle H_{j} \rangle \,,
\label{eq:U}
\end{equation}
is not a symmetry of the Lagrangian does CP violation occur. If $U$ is a symmetry of the Lagrangian, CP is conserved even though the VEVs are complex \,\cite{Branco:1983tn}.

\section{Geometrical CP Violation \label{sec:GCPV}}

Here we discuss the particular case where the phases that appear in the scalar VEVs are determined independently of the arbitrary parameters of the scalar potential. These are referred to as calculable phases, and the framework as geometrical CP violation (GCPV). GCPV requires at least three Higgs doublets and a non-Abelian symmetry \,\cite{Branco:1983tn}. $\Delta(27)$ (a discrete subgroup of $SU(3)$) was the first group found to produce such calculable phases \,\cite{Branco:1983tn}. In \,\cite{deMedeirosVarzielas:2011zw} this was generalised to larger groups obtaining the same calculable phases and more recently, several new phase solutions were advanced and expressed in terms of the number of scalars and the group \,\cite{Varzielas:2012pd}. We will briefly review the framework, explaining the origin and clarifying some of the possible generalisations in some detail. The notation follows mostly from \,\cite{Varzielas:2012pd}, with $\eta_N \equiv e^{2\text{i}\pi/N}$ and $\langle H_i \rangle \equiv v e^{\text{i} \alpha_i}$ (the assumption that the VEVs have the same magnitude is justified by the respective potentials favouring this type of minima in very general cases).

The only phase dependence in the renormalisable $\Delta(27)$ potential for a scalar triplet can be written as
\begin{equation}
H_1^2 (H_2 H_3)^\dagger + c.p.= v^4 (e^{\text{i} A_1} + e^{\text{i} A_2} + e^{\text{i} A_3}) \,,
\label{eq:leading3}
\end{equation}
where $c.p.$ represents the cyclic permutations and the phases $A_i$ are $A_1=2\alpha_1 - \alpha_2 - \alpha_3$ and permuting, $A_2=-\alpha_1 + 2\alpha_2 - \alpha_3$,  $A_3=-\alpha_1 - \alpha_2 + 2 \alpha_3$. Due to a combination of the cyclic properties of the group and invariance under the gauge group, $\sum A_i=0$ must be verified \,\cite{Varzielas:2012pd}.
If we constrain ourselves to renormalisable potentials this invariant is the only phase dependence in the potential (with its hermitian conjugate, $h.c.$), and the minimum of the scalar potential depends on
\begin{equation}
V_3 =  (e^{\text{i} A_1} + e^{\text{i} A_2} + e^{\text{i} A_3}) + h.c. \,,
\end{equation}
If $V_3$ appears in the potential with negative coefficient, $V_3$ should be maximised corresponding to a contribution of $+6$ for $A_i=0$. Otherwise, $V_3$ should be minimised by the phases - but a contribution of $-6$ is not allowed, as $A_i= \pi$ would violate $\sum A_i=0$.
\begin{figure}
	\centering
		\includegraphics[width=3 cm,keepaspectratio=true]{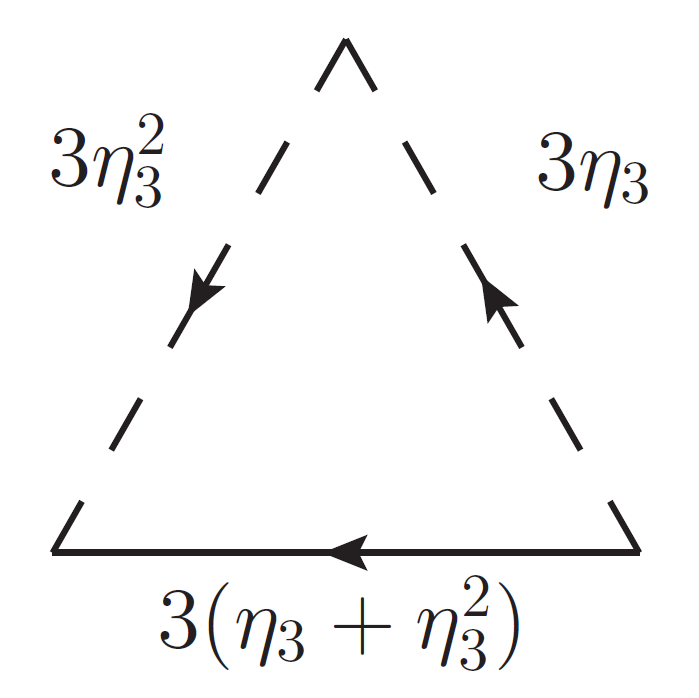}
	\caption{\label{fig:triangle} The equilateral triangle shape associated with the minimising solution for $V_3$.}
\end{figure}
Fig. \ref{fig:triangle} illustrates how the minima with $A_i = \pm 2 \pi/3$ corresponds to a contribution of $-3$.
The $\Delta(27)$ GCPV VEVs like $(\eta_3^{\mp 1},1,1)$ (corresponding to $\alpha_1= \mp 2 \pi / 3$, $\alpha_2=\alpha_3=0$) are obtained from such configurations. The same solutions can be obtained in larger groups containing $\Delta(27)$ \,\cite{deMedeirosVarzielas:2011zw}.

\begin{figure}
	\centering
		\includegraphics[width=5 cm,keepaspectratio=true]{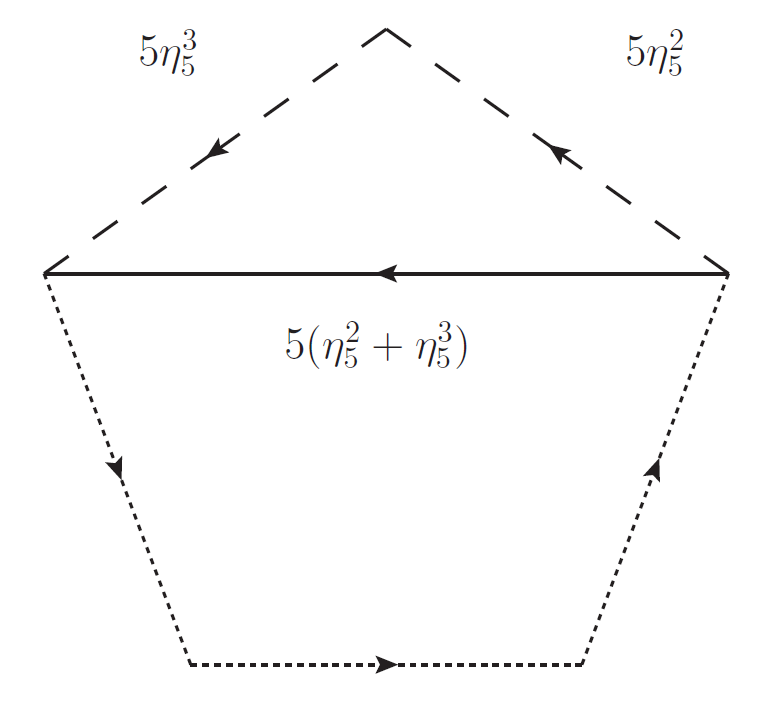}
	\caption{\label{fig:pentagon} The pentagon shape associated with the minimising solution for $V_5$.}
\end{figure}

$\Delta(27)$ is the semi-direct product of cyclic groups $C_3 \ltimes (C_3 \times C_3)$, and by considering analogous semi-direct products it is possible to find new GCPV candidates - although in general only by going to non-renormalisable potentials. A direct generalisation involves groups $C_N \ltimes (C_N \times ... \times C_N)$ with $N-1$ factors of $C_N$ inside the brackets. The properties of the group lead to the lowest order phase dependent invariant being a direct generalisation of Eq.(\ref{eq:leading3})
\begin{equation}
H_1^{N-1} \left( H_2 (...) H_N \right)^\dagger + c.p. \,.
\label{eq:leadingN}
\end{equation}
The order of such invariants is $2(N-1)$ and already for $N=4$ this is of order 6 and non-renormalisable.
For even number of scalars $N$ the situation is arguably less interesting, as $A_i= \pi$ is an allowed solution to minimise the invariant (for $N$ even it verifies $\sum A_i=0$ as required). But with an odd number of scalars $N$ one has e.g. for $N=5$
\begin{equation}
V_5 =  (e^{\text{i} A_1} + e^{\text{i} A_2} + e^{\text{i} A_3}+e^{\text{i} A_4}+e^{\text{i} A_5}) + h.c. \,,
\end{equation}
in this particular case with $A_1=4\alpha_1 - \alpha_2 - \alpha_3 - \alpha_4 - \alpha_5$. As was the case with $N=3$, $A_i= \pi$ is not allowed due to the requirement $\sum A_i=0$.
It turns out that to minimise the generalised $V_N$ the requirement is \,\cite{Varzielas:2012pd}
\begin{equation}
A_i= \pm \frac{N-1}{2} \frac{2 \pi}{N} \,,
\label{eq:oddNmin}
\end{equation}
corresponding to adjacent sides of a regular $N$-sided polygon, as seen in Fig. \ref{fig:pentagon} for $N=5$.
This can be obtained e.g. with
\begin{equation}
\alpha_1 = \mp \frac{N-1}{2} \frac{2 \pi}{N} \,,
\label{eq:oddNalpha}
\end{equation}
and the remaining phases vanishing. The VEVs are then of the type
\begin{equation}
 v \left(\eta_N^{\mp(N-1)/2}, 1,(...),1 \right) \,.
\label{eq:oddNVEV}
\end{equation}

If we consider instead the more general groups $C_N \ltimes (C_n \times ... \times C_n)$ with $N-1$ factors of $C_n$ inside the brackets, there are some cases of interest: (1) $n=kN$ (an integer multiple); (2) $n$ is not a multiple but shares a prime factor with $N$; (3) $n$ and $N$ share no prime factors. The first case is the most straightforward - depending on the irreducible representation (irrep) chosen one can obtain the same GCPV VEVs that are available with $k=1$, or alternatively the choice of irrep forbids the invariant in Eq.(\ref{eq:leadingN}), leaving as lowest order phase dependent invariant
\begin{equation}
\left( H_1^{N-1} \left( H_2 (...) H_N \right)^\dagger \right)^k + c.p. \,,
\label{eq:leadingkN}
\end{equation}
of order $2k(N-1)$. The solutions to Eq.(\ref{eq:oddNmin}) for odd $N$ are then changed from Eq.(\ref{eq:oddNVEV}) to \,\cite{Varzielas:2012pd}
\begin{equation}
v \left( \eta_{kN}^{\mp \frac{(2l-1)N-1}{2}},1,(...),1 \right) \quad \text{for} \, l \leq k  \,.
\label{eq:oddNkVEV}
\end{equation}
For a discussion of the second case (which always applies when $n$ and $N$ are both even) we refer the interested reader to \,\cite{Varzielas:2012pd}. The third case merits some detailed discussion here - due to invariance under the gauge group and as there is no shared prime factor, the relevant phase dependent invariant is
\begin{equation}
\left( H_1 H_2^\dagger \right)^n + c.p. \,,
\label{eq:leadingn}
\end{equation}
which differs from the previously featured invariants in an important way: each individual term no longer has all the $H_i$ in it. For $N>3$ this means that there can be independent invariants 
$\left( H_1 H_3^\dagger \right)^n + c.p.$, and $\left( H_1 H_4^\dagger \right)^n + c.p. $ and so on. For odd $N$ there are  $(N-1)/2$ independent invariants when the $h.c.$ is considered, for $N$ even there is one unpaired invariant that is its own $h.c.$ so the total is $N/2$. It is interesting that for $n=2$ this class of invariants appears at order 4 and they are therefore renormalisable. In the case $N=3$ and $n=2$ (which corresponds to the group $\mathbf{A}_4$), if we want to minimise the phase-dependent invariant $I_{32}=\left( H_1 H_2^\dagger \right)^2 + \left( H_2 H_3^\dagger \right)^2+ \left( H_3 H_1^\dagger \right)^2 +h.c.$, then the phases $\alpha_1=0$, $\alpha_2=-2\pi/6$ and $\alpha_3=-4\pi/6$, of VEV $(1,\eta_6^{-1},\eta_6^{-2})$, result in the $A_i=2\pi/3$ solution depicted in Fig. \ref{fig:triangle}, but note the respective $U$ from Eq.(\ref{eq:U}) leaves $I_{32}$ invariant apart from a global phase. Indeed this is the case for any $n$ and odd $N$, the invariant $I_{Nn}=\left( H_1 H_2^\dagger \right)^n+c.p. +h.c.$ is minimised by $n\alpha_i=(-i+1)\frac{N-1}{2} \frac{2 \pi}{N}$ or by $n\alpha_i=(i-1)\frac{N-1}{2} \frac{2 \pi}{N}$ (starting from $n\alpha_1=0$). This can be obtained by solving the equations $n\alpha_1-n\alpha_2=\pm \frac{N-1}{2} \frac{2 \pi}{N}$ or equivalently by picturing the regular polygon with $N$ sides, and assigning consecutive $n\alpha_i$ either clockwise or anti-clockwise always leaving $(N-1)/2$ sides between consecutive $n\alpha_i$ (and also between the last $n\alpha_N$ and $n\alpha_1=0$). In any case, the $U$ relating such a solution to its conjugate only shifts each term of the invariant $I_{Nn}$ by the global phase $\pm 2n\alpha_N$ (the $h.c.$ is shifted by $\mp 2n\alpha_N$).
In addition, for odd $N>3$, there are multiple invariants appearing at order $2n$ with different phase dependences, say $A_i$, $B_i$ etc. It is not possible to have $A_i=\pm \frac{N-1}{2} \frac{2 \pi}{N}$ simultaneously with $B_i=\pm \frac{N-1}{2} \frac{2 \pi}{N}$ (and not with all $B_i=0$ either). This is the case already for $N=5$, where one can take $A_1=n(\alpha_1 - \alpha_2)$ and $B_1=n(\alpha_1 - \alpha_3)$, with the respective invariants being  renormalisable only for $n=2$. As noted above, $\alpha_i=(-i+1)4\pi/5n$ minimises the $A_i$ invariant (which corresponds to $I_{5n}$) with all $A_i=4 \pi/5$. But this type of phase solution simultaneously corresponds to all $B_i= 8\pi/5$ which does not extremise the $B_i$ invariant (indeed each $B_i$ can be expressed as a sum of two $A_i$, e.g. $B_1=A_1+A_2$).
For even $N$ there are some differences: while $A_i=\pi$ may not be compatible with $B_i=\pi$, it is compatible with $B_i=0$. For $N=4$, $A_1=n(\alpha_1-\alpha_2)$, and from the self-conjugate invariant the dependence is $B_1=n(\alpha_1-\alpha_3)$; then $A_i=\pi$ can be obtained from e.g. $\alpha_{1,3}=\pi/2n=-\alpha_{2,4}$ which also gives $B_i=0$ which is the desired minimising solution if the $B_i$ invariant has a negative coefficient. For $N=6$, $A_1=n(\alpha_1-\alpha_2)$, $B_1=n(\alpha_1-\alpha_3)$, and from the self-conjugate invariant the dependence is $C_1=n(\alpha_1-\alpha_4)$; then $A_i=C_i=\pi$ and $B_i=0$ are obtained for e.g. $\alpha_{1,3,5}=\pi/2n=-\alpha_{2,4,6}$ and this is a minimising solution as long as the coefficients of the $A_i$, $B_i$ and $C_i$ invariants are respectively positive, negative and positive. Once again, the $U$ associated to the solution only shifts the invariant by a global phase.

\section{Fermion mixing with geometrical CP violation}

When fermions are added to the framework of GCPV it is not trivial to construct viable models. Leading order structures were suggested in \,\cite{deMedeirosVarzielas:2011zw}. Calculable phases are also robust when the potential includes non-renormalisable terms \,\cite{Varzielas:2012nn}. Based on these results, a minimal model of GCPV fitting all data was proposed in \cite{Bhattacharyya:2012pi}. The group employed is $\Delta(27)$ and a minimal amount of additional matter is added.

The three Higgs doublets, $H_i$, transform as a $\Delta(27)$ triplet assigned to a $\mathbf{3}_{01}$ irrep (in this section we denote them with a lower index). Their hermitian conjugates $H^{\dagger i}$ transform as the conjugate representation $\mathbf{3}_{02}$ (in this section we denote them with an upper index). The relevant generators of the group will be denoted as $c$ (cyclic permutation) and $d$ (diagonal phases). They act on the irreps as $c (H_1, H_2, H_3) \rightarrow (H_2, H_3, H_1)$, $c (H^{\dagger 1}, H^{\dagger 2}, H^{\dagger 3}) \rightarrow (H^{\dagger 2}, H^{\dagger 3}, H^{\dagger 1})$, and $d (H_1, H_2, H_3) \rightarrow (H_1, \omega H_2, \omega^2 H_3)$, $d (H^{\dagger 1}, H^{\dagger 2}, H^{\dagger 3}) \rightarrow (H^{\dagger 1}, \omega^2 H^{\dagger 2}, \omega H^{\dagger 3})$. $\omega \equiv \mathrm{e}^{\mathrm{i} 2 \pi/3}$ corresponds to the $\eta_3$ used in section \ref{sec:GCPV}.
There are nine singlet irreps $\mathbf{1}_{ij}$, where the subscript $\{ij\}$ denotes how they transform under the generators $c \mathbf{1}_{ij} = \omega^i \mathbf{1}_{ij}$, $d \mathbf{1}_{ij} = \omega^j \mathbf{1}_{ij}$. Further details about $\Delta(27)$ can be found in the references of \cite{Bhattacharyya:2012pi}.

\begin{figure}
\begin{center}
 \includegraphics{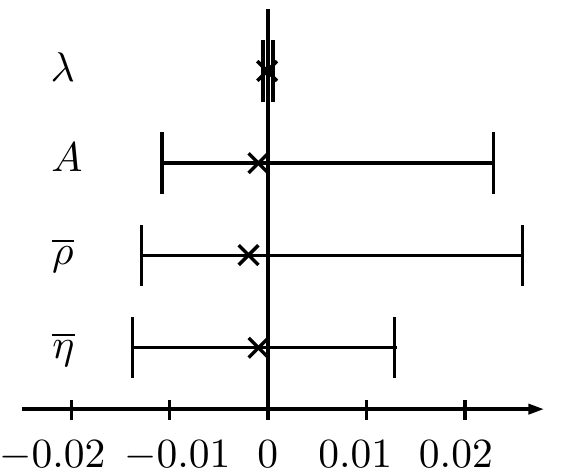}
 \caption{The experimental spread of the Wolfenstein parameters $\lambda$, $A$, $\bar{\eta}$ and $\bar{\rho}$ around their central values. Crosses denote our model values. \label{fig:Wolfenstein}}
\end{center}
\end{figure}

From the solution in Fig. \ref{fig:triangle} we can have a complex VEV of the type:
\begin{equation}
\langle H_i \rangle = v(\omega,1,1) \,.
\label{eq:VEV}
\end{equation}
This VEV necessarily violates CP, as the corresponding $U$ (see Eq.~(\ref{eq:U})) does not leave the potential invariant. 
The full scalar potential is presented later as we first focus on the Yukawa interactions of the quarks.
The possible Yukawa contractions in this framework were considered briefly in \,\cite{Branco:1983tn} and in more detail in \,\cite{deMedeirosVarzielas:2011zw}:
in order to make invariant Yukawa terms some of the quarks must transform as triplet or anti-triplet under $\Delta(27)$. We write the invariants symbolically as $Q H_i d^c$ and $Q H^{\dagger i} u^c$ (without explicit $SU(2)$ indices). $Q$ are the left-handed quark doublets and $u^c$, $d^c$ are the up and down right-handed singlets. If $Q_i$ is a $\mathbf{3}_{01}$ we would necessarily require the $d^c_i$ to transform also as a $\mathbf{3}_{01}$, conversely if $Q^i$ is a $\mathbf{3}_{02}$, $u^{c i}$ is forced to be a $\mathbf{3}_{02}$. At least one sector has a leading order Yukawa structure given by the $\Delta(27)$ invariant $\mathbf{3}_{0i}\otimes\mathbf{3}_{0i}\otimes\mathbf{3}_{0i}$. With the symmetrical VEV in Eq.~(\ref{eq:VEV}), this structure leads to a mass matrix with degenerate quark masses \,\cite{Branco:1983tn} and so the conclusion is that $Q$ cannot be assigned as a triplet or an anti-triplet.
We choose instead $u^c$ and $d^c$ as $\Delta(27)$ triplets so that $Q H_i d^{c j}$ and $Q H^{\dagger i} u^c_j$ are invariant with $Q$ as singlets. Both sectors then have Yukawas from the $\Delta(27)$ invariants $\mathbf{1}_{ij} \otimes (\mathbf{3}_{01}\otimes\mathbf{3}_{02})$ \,\cite{deMedeirosVarzielas:2011zw}.
Although $\mathbf{3}_{01}\otimes\mathbf{3}_{02}$ results in 9 distinct singlets, the group properties are such that any $\mathbf{3}_{01}\otimes\mathbf{3}_{02} \rightarrow \mathbf{1}_{ij}$ with $i\neq0$ explicitly involves powers of $\omega$ (complex), so not all possibilities are allowed by CP invariance of the Lagrangian. To have renormalisable Yukawa interactions we assign $Q_1$, $Q_2$ and $Q_3$ each as one of the three $\mathbf{1}_{0i}$ singlets. The possibilities are assigning all three $Q$ in the same singlet irrep, or assigning two in the same, or all three $Q$ in different irreps. All three structures lead to mass matrices that have a special structure defined by rows. When $Q$ is a $\mathbf{1}_{00}$, $\mathbf{1}_{01}$ or $\mathbf{1}_{02}$, the respective $H_i d^{c j}$ or $H^{\dagger i} u^{c}_j$ product is $\mathbf{1}_{00}$, $\mathbf{1}_{02}$ or $\mathbf{1}_{01}$ respectively, which essentially amounts to a shift in the position of the $\omega$ in the mass matrix. More explicitly the corresponding down mass matrix looks like:

\begin{equation}
\tilde{M}_d = v \begin{pmatrix}
	y_{1} \omega & y_{1} & y_{1} \\
	y_{2} & y_{2} \omega & y_{2} \\
	y_{3} & y_{3} & y_{3} \omega 
\end{pmatrix}
\end{equation}
and the associated up quark mass matrix looks very similar with $\omega^2$ instead of $\omega$ and the second and third rows swapped.
If instead $Q_1$, $Q_2$, and $Q_3$ are assigned to $\mathbf{1}_{00}$, $\mathbf{1}_{00}$, and $\mathbf{1}_{02}$ respectively, we get:
\begin{equation}
M_d = v \begin{pmatrix}
	y_{1} \omega & y_{1} & y_{1} \\
	y_{2} \omega & y_{2} & y_{2} \\
	y_{3} & y_{3} & y_{3} \omega 
\end{pmatrix}
\end{equation}
We recall that due to the explicit CP invariance of the Lagrangian the couplings are real and the phase appears only through the VEV.
Consider then the hermitian matrices $M M^\dagger$:

\begin{equation}
\tilde{M}_d \tilde{M}_d^{\dagger}= 3 v^2 \begin{pmatrix}
	y_{1}^2 & 0 & 0 \\
	0 & y_{2}^2 & 0 \\
	0 & 0 & y_{3}^2
\end{pmatrix}
\end{equation}
\begin{equation}
M_d M_d^{\dagger}= 3 v^2 \begin{pmatrix}
	y_{1}^2 & y_{1} y_{2} & 0 \\
	y_{1} y_{2} & y_{2}^2 & 0 \\
	0 & 0 & y_{3}^2
\end{pmatrix}
\end{equation}
The vanishing off-diagonal entries are a consequence of $1+\omega+\omega^2=0$.
The determinant of $M_d M_d^{\dagger}$ is zero but it has two non-vanishing masses, and the choice with all generations of $Q$ in the same singlet irrep (not shown) leads to a rank 1 structure with a single non-vanishing mass. It is apparent that these hermitian structures are real.
To obtain a viable CKM matrix we need to generate additional off-diagonal terms and to somehow preserve the complex phase. Off-diagonal terms appear in $M M^\dagger$ when rows have the $\omega$ in the same column, and the minimal way to get them is to add a gauge singlet scalar  $\phi$ that is a non-trivial $\Delta(27)$ singlet. Without loss of generality we place $\phi$ in the irrep $\mathbf{1}_{01}$, enabling a new non-renormalisable Yukawa coefficient populating each row from $Q H_i d^{c j} \phi$. For $Q_1$, $Q_2$, and $Q_3$ in $\mathbf{1}_{00}$, $\mathbf{1}_{00}$, and $\mathbf{1}_{02}$ respectively, we add to $M_d$ the corresponding matrix:
\begin{equation}
M_{\phi} = v \begin{pmatrix}
	y_{\phi1} & y_{\phi1} \omega & y_{\phi1} \\
	y_{\phi2} & y_{\phi2} \omega & y_{\phi2} \\
	y_{\phi3} \omega & y_{\phi3} & y_{\phi3} 
\end{pmatrix}
\label{eq:yphi}
\end{equation}
From the interference $M_{d} M_{\phi}^\dagger + M_{\phi} M_{d}^\dagger$ we obtain the required additional off-diagonal entries. The effect of $M_{\phi} M_{\phi}^\dagger$ can be absorbed within the existing structure of $M_d M_{d}^{\dagger}$. 
We now want a complex phase in the CKM matrix, from complex $M M^\dagger$. The minimal possibility is to consider non-renormalisable interactions with higher powers of $H$ e.g.~$Q H_i d^{c j} (H_k H^{\dagger l})$. It turns out that the non-trivial structure extracted from such interactions is
\begin{equation}
M_{H} = v \begin{pmatrix}
	y_{H1}  & y_{H1} \omega^2 & y_{H1} \omega^2 \\
	y_{H2} & y_{H2} \omega^2 & y_{H2} \omega^2 \\
	y_{H3} \omega^2 & y_{H3} \omega^2 & y_{H3}
\end{pmatrix}
\end{equation}
where the identity $1+\omega+\omega^2=0$ was used and the existing coefficients were redefined to absorb similar entries in the mass matrix. From the interference $M_{d} M_{H}^\dagger + M_{H} M_{d}^\dagger$ we obtain phases in $M M^\dagger$ that enable complex CKM elements, $M_{\phi} M_{H}^\dagger + M_{H} M_{\phi}^\dagger$ and $M_{H} M_H^\dagger$ give structures that do not qualitatively change the analysis.
$M_{\phi}$ and $M_{H}$ are the minimal mandatory additions needed for a perfect fit to the existing data.
The Lagrangian (showing the $\Delta(27)$ indices) is:
\begin{equation}
\mathcal{L} = Q \left(H^{\dagger i} u^{c}_j + H_i d^{cj} + H_i d^{cj} \phi + H_i d^{cj} (H_k H^{\dagger l})  \right) \,.
\end{equation}
We found that the only choice that favourably accounts for the precision flavour data is $Q_1$, $Q_2$ and $Q_3$ as $\mathbf{1}_{00}$, $\mathbf{1}_{00}$ and $\mathbf{1}_{02}$ respectively. In the up quark sector, $M_u M_u^\dagger$ can be considered diagonal, and we need only one additional non-renormalisable Yukawa in order to generate the small up quark mass (recall the determinant of the renormalisable structure is zero for this choice of irreps). In Fig.~\ref{fig:Wolfenstein} we show that this choice can successfully reproduce the Wolfenstein parameters, and one can also compare the model values (in the right column) with the experimental values:
\begin{equation}
\label{eq:wolfenstein}
\begin{aligned}
 \lambda^\text{exp} &= 0.22535 \pm 0.00065 & \lambda &= 0.22534, \\
 A^\text{exp} &= 0.811 \begin{matrix}
 +0.022\\
 -0.012
 \end{matrix} & A &= 0.810,\\
 \bar{\rho}^\text{\;exp} &= 0.131 \begin{matrix}
  +0.026\\
  -0.013
 \end{matrix} & \bar{\rho}  &= 0.129,\\
 \bar{\eta}^\text{\;exp} &= 0.345 \begin{matrix}
 +0.013\\
 -0.014
 \end{matrix} & \bar{\eta} &= 0.344.
\end{aligned}
\end{equation}

We present now the full scalar potential. It contains the $\Delta(27)$ triplet $H_i$ but also $\phi$ which we introduced to obtain desirable Yukawa structures:
\begin{align}
 V(H,\phi) =& m_1^2 \left[ H_1 H_1^\dagger \right]+
m_2^2 \phi \phi^\dagger + 
m_3 (\phi^3 + h.c.)\\
&+\lambda_1 \left[(H_1 H_1^\dagger)^2\right]+
\lambda_2 \left[H_1 H_1^\dagger H_2 H_2^\dagger \right]+
\lambda_3 \left[H_1 H_2^\dagger H_1 H_3^\dagger
+h.c.\right]\\
&+\lambda_4 (\phi\phi^\dagger)^2+
\lambda_5 \left[\phi (H_1 H_2^\dagger)+h.c.\right]+
\lambda_6 \left[\phi \phi (H_1 H_3^\dagger)+ h.c. \right]\,,
\end{align}
where the coefficients are real, and the square brackets represent also the cyclic permutations on the $\Delta(27)$ indices which we do not explicitly show.
While the geometrical phase solution in Eq.~(\ref{eq:VEV}) is not affected by $\phi\phi^\dagger$, it is only when $\lambda_5$ and $\lambda_6$ are small that it holds. One can add a $Z_4$ symmetry acting on $\phi$ to trivially enforce these couplings to vanish (in this case Eq.(\ref{eq:yphi}) arises from a $\phi^4$ insertion instead of $\phi$, all conclusions remaining unchanged).

Following the minimisation of the potential and determination of the mass eigenvalues, we observed these features (for illustration we display only the CP-even scalar components): $(i)$ The $\phi$ field is much heavier (beyond $1$\,TeV) and decouples from the $SU(2)$ doublets. More specifically, the mass of $\phi$ is determined by $\lambda_{\{4,5,6\}}$, while those of $h_{\{a,b,c\}}$ are controlled by $\lambda_{\{1,2,3\}}$.
$(ii)$ The physical scalars $h_a, h_b$ and $h_c$ mix in a very specific way (see \cite{Bhattacharyya:2012pi} and references therein): the scalar mass squared matrix having the structure
\begin{equation}
	\begin{pmatrix}
		A & B & B\\
		B & C & D\\
		B & D & C
	\end{pmatrix},
\end{equation}
leading to one physical scalar $h_a$ that is orthogonal to the other scalars and having no $h_a VV$-type gauge couplings ($V=W,Z$). The Yukawa couplings of $h_a$ to up- and down-type quarks are strongly suppressed except for the $h_a ct$ and $h_a uc$ couplings which are about $0.45$. The other physical scalars, $h_b$ and $h_c$, have nearly SM-like gauge and Yukawa couplings.
 
\begin{figure}
\begin{center}
 \includegraphics{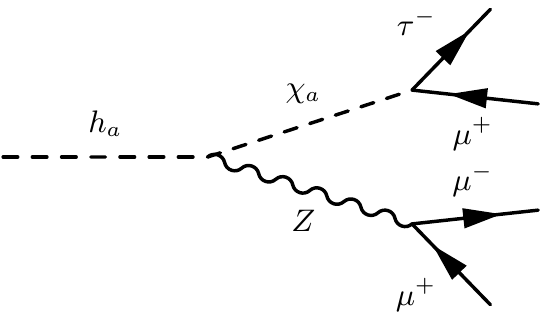}
 \caption{Example of a decay mode of the exotic scalar $h_a$ that can be tested at the LHC. \label{fig:Feynman}}
\end{center}
\end{figure}

Depending on the scalar potential couplings, two viable scenarios are identified: (I) There is only one light scalar, $h_b$, assuming the role of the SM-like Higgs found near $125$\,GeV, with all other scalars beyond the current exclusion range of the LHC. This is a decoupling limit which reproduces almost SM-like scalar structure.  
(II) A scenario with richer collider consequences is possible when the exotic scalar $h_a$ is light enough to be produced at the LHC e.g. through $h_a uc$ or decays of tops or heavy scalars. Under the reasonable assumption that $m_\phi >1$\,TeV or so we can obtain the analytic relations:
  \begin{align}
  m_{h_a}^2 &= \frac{2}{3} \left(2 \lambda_1 v^2-2 \lambda_2 v^2+3 \lambda_3 v^2\right),\\
  m_{h_{c/b}}^2 &= \frac{1}{6} \bigl(5 \lambda_1 v^2+4 \lambda_2 v^2
  \pm\sqrt{3} \bigl[v^4 \bigl(3 \lambda_1^2+8 \lambda_1
   \lambda_2\nonumber\\&\quad-16 \lambda_1 \lambda_3+16 \lambda_2^2-64 \lambda_2
   \lambda_3+64 \lambda_3^2\bigr)\bigr]^\frac{1}{2} \bigr).
  \end{align}
The $\lambda_i$ can be adjusted to give $m_{h_a}$ around the mass $125$\,GeV of the SM-like $h_b$, with $h_c$ heavier than $600$\,GeV.
In this case, there can be a spectacular decay channel through $h_a\to\chi_a Z$, if $m_{\chi_a} \sim 20$\,GeV, with the pseudo-scalar $\chi_a $ then decaying to charged leptons of different flavours (e.g. $\mu \tau$) and the $Z$ boson decaying to leptons, as in Fig.~(\ref{fig:Feynman}). There is enough freedom in the lepton sector to boost coupling of $\chi_a$, which may have a sizeable branching ratio in this channel, but a more specific prediction requires a detailed numerical study of the lepton Yukawa sector.

Although in both scenarios (I) and (II) there are a few heavy scalars above the current LHC limit of  650 GeV or so, their relative heaviness compared to the SM Higgs is not a result of fine-tuning of parameters as for each heavy state there is a reasonably independent combination of $\lambda_i$-type couplings which simply has to be set to a higher value (we have verified this numerically).
If the LHC bound goes up, we have to accordingly raise the maximum allowed value of some $\lambda_i$ beyond $\pi$ e.g. up to maximum allowed value of $2 \pi$ to go over $1$\,TeV.
A final consequence worth pointing out is that the additional scalar states below $1$\,TeV in both scenarios (I) and (II), all coupling to SM gauge bosons, would affect the energy dependence of longitudinal gauge boson scattering. This energy dependence might be different from the SM expectation due to the extra scalars, whose quantitative impact may be probed at the high luminosity option of the LHC depending on their masses and couplings.

\section{Summary}

We have reviewed and clarified the framework of geometrical CP violation in the context of multi-Higgs models.
We then reviewed the realisation of spontaneous CP violation of geometrical origin in a minimal $\Delta(27)$ flavour model that reproduces the CKM mixing matrix. This scenario is quite falsifiable, as only two choices broadly worked, out of which only one set of representations fits the precision of flavour data. The scalar sector of this model retains some symmetry of the flavour group and this can lead to exotic scalar decays into multi-lepton of different flavours, which would be a smoking gun signal of the model that is testable at the LHC.

\ack
The author thanks CFTP for hospitality and also the organisers of Discrete 2012 for a very nice conference.
The author was supported by DFG grant PA 803/6-1 and partially through PTDC/FIS/098188/2008.


\end{document}